\documentstyle[12pt]{article}
\def\dspace{\baselineskip=0.3 in}
\begin{document}
\dspace

\centerline{\bf DUAL NATURE OF THE RICCI SCALAR, CREATION OF}

\centerline{\bf SPINLESS AND SPIN - 1/2 PARTICLES AS WELL AS}

\centerline{\bf A NEW COSMOLOGICAL SCENARIO I}
\vspace{1cm}

\centerline{\bf S.K.Srivastava,}
\centerline{\bf Department of Mathematics,}

\centerline{\bf North Eastern Hill University,}

\centerline{\bf NEHU Campus,}

\centerline{\bf Shillong - 793022}

\centerline{\bf ( INDIA )}

\centerline{\bf e - mail : srivastava@nehu.ac.in}

\vspace{2cm}

In earlier papers [11-17], it is found that  the Ricci scalar
behaves in a dual manner (i) like a matter field and (ii)
like a geometrical field. In ref.[17], inhomogeneous  cosmological models are derived using dual
roles of the Ricci scalar. The essential features of these models is
capability of these to exhibit gravitational effect of compact
objects also in an expanding universe. Here, production of spinless and spin-1/2 particles are demonstrated in these models.



\newpage

\centerline{\bf 1.Introduction}

\bigskip

  Consideration of higher-derivative gravity is not new and it is a strong
  candidate in the context of string theories and Einstein-Gauss-Bonnet
  gravity. In the context of the early universe,  use of higher-derivative
  gravity  started from more than  three decades ago. In 1977, Stelle
  considered it and pointed out that, in contrast to square of Weyl tensor,
  introduction of $R^2$ or $f(R)$ terms did not lead to the ghost problem
  [1]. Here $R$ is the Ricci scalar and $f(R)$ is the linear combination of
  $R^2$ terms subject to the condition $lim_{R\to 0} f(R)/R = 0$ [2]. From
  these theories, it is found that the Ricci scalar behaves as a physical
  field also [3-10].

While quantizing gravity (quantizing
components of the metric tensor), this theory has
problem at the perturbation level where ghost terms appear in the Feynman
propagator of gravitons. Around 25 years back, Starobinsky noted that if
coupling constants in the action of higher-derivative gravity are taken
properly, ghost terms do not appear and only one  scalar particle, represented
by the Ricci scalar, is obtained with positive energy as well as positive
squared mass [6,7]. He called this particle as ``scalaron''[6,7]. In papers [1-8],gravitational constant $G$ is either taken equal to unity or lagrangian density is taken as $\frac{1}{16 \pi G} (R +$ higher - derivative terms). As a result, (mass)$^2$ of the Ricci scalar, does not depend on the gravitational constant $G$. As $G$ has a very important role in gravity, in papers [11 - 17] as well as here, lagrangian density is taken as $(\frac{1}{16 \pi G} R +$ higher - derivative terms) leading to a drastic change where (mass)$^2$ of $R$ depends on $G$ also. In these papers [11 - 17], choosing coupling constants of the higher-derivative gravitational action in  a suitable manner, it is found
that  the Ricci scalar
behaves like a physical particle,  called as `` riccion'', with positive (mass)$^2$ without 
any ghost problem. This  feature
leads to manifestation of Ricci scalar in dual manner (like a geometrical
field as well as a scalar particle). It is important to mention that riccion
is a scalar particle (having only one mode unlike a graviton having five modes), but it is different from scalar mode of graviton. 
Detailed discussion on this difference is given in Appendix A. Riccion is
given by a scalar field ${\tilde R} = \eta R,$ where $\eta$ is a constant
having length dimension. A ``riccion'' is different from a ``scalaron'' in two
ways (i) mass dimension of riccion is one like other scalar fields used in a
physical theory, such as quintessence, inflaton and Higg's particle, whereas
mass dimension of scalaron is two and (ii) $(mass)^2$ of scalaron does not
depend on $G$, whereas riccion $(mass)^2$ depends on $G$.

In refs.[11-17], some interesting 
results are obtained using dual role of the Ricci scalar. One of
 the
important consequences of the same is derivation of models of the
 early
universe using a physical method of phase transition and spontaneous
symmetry breaking. In ref.[11], spatially homogeneous cosmological
 models
were derived 
using these methods and it was shown that universe should bounce at
the
critical temperature $\sim 10^{18}$GeV. In ref.[15], employing these
methods,
models of the early universe are
derived exhibiting inhomogeneity and anisotropy at small scales.It is found that
these
models are asymptotically homogeneous and isotropic.

The theory of gravity is manifested through geometry of the
 space-time.
Two types of space-times are used for manifestation of 
gravitational interactions. 
One type is static space-times representing geometry around compact
objects e.g., Schwarzschildspace-time,
 Riessner-Nordstr$\ddot {\rm o}$m
space-time and others. The other type of space-times are homogeneous 
and
inhomogeneous models representing the expanding universe. Former 
type of models
 do not give any information about the expanding universe and the
latter type of models are unable to account for gravitational effect of
 individual
objects in the expanding universe. So, models  exhibiting local
gravitational effect of an individual object as well as expansion
of the universe  simultaneously are required to discuss evolution of
the universe in a more natural way. Using dual roles of the Ricci
scalar, such cosmological models are
derived in the reference [17], giving the first two stages of the
beginning of the early universe.

In the present paper, production of spinless and spin-1/2 particles is demonstrated in the inhomogeneous models derived in the reference [17] . Contributions of these particles to the development of the universe will be discussed in paper II.

The paper is organized as follows. For convenience of the reader, a review of reference[17] is given in section 2.

In section 3, taking interaction of riccions with another scalar field $\Phi$,
production of spinless particles is demonstrated. In section 4, creation of
spin-1/2 particles has been shown. Cosmological consequences of these created
particles are planned to be discussed in the second paper of the series.

Natural units ($\hbar = c = 1$) are used throughout the paper. Here
$\hbar$ and $c$ have their usual meaning. For the sake of convenience, it is
given that $1 {\rm GeV} = 1.16 \times 10^{13}K  = 1.78 \times 10^{- 24}$
g, 1 GeV$^{-1}$ = $1.97 \times 10^{- 14}$ cm = $6.58 \times 10^{-25}$ Sec.

\bigskip

\noindent \underline {\bf  2.(a) Dual role of the Ricci scalar and riccion}

\bigskip

In  this section, a brief review of the reference [17] is given. Here, the gravitational action with thermal 
correction is taken as 
\begin{eqnarray*}
 S_g &=& \int {d^4x}{\sqrt{- g}}\Big[ -
{\frac{R}{16{\pi}G }} + 4{\lambda}{\alpha} T^2 R  +
{\alpha}R^2 \\ && - 16{\eta}^2 {\lambda}{\alpha}(R^3 +
9 {\Box} R^2)  + \frac{\eta^2}{48} R^3 \Big] ,
\end{eqnarray*}

$$  \eqno(2.1)$$
where   ${\alpha}$ and ${\lambda}$ are
 dimensionless
coupling constants. Here $T$ is the temperature, $G$ is the gravitational constant and 
${\eta}$ is another constant with 
length dimension, which is  used for dimensional corrections only [11-17] .
The invariance of $S_g$ , given by eq.(2.1) , under  
transformations
$g_{\mu\nu}\to g_{\mu\nu} 
+{\delta}g_{\mu\nu}$ yields the gravitational field equations

$$\Big[ -
{\frac{1}{16{\pi}G }} + 4{\lambda}{\alpha} T^2\Big](R_{\mu\nu} -{\frac{1}{2}}g_{\mu\nu}R) + {\alpha} (2 R_{;\mu\nu} - 2 g_{\mu\nu}{\Box} R  $$ $$- {\frac{1}{2}} g_{\mu\nu} R^2 + 
2 R R_{\mu\nu}) - 16{\eta}^2 {\lambda}{\alpha}\{3 R^2_{; \mu\nu} - 3 g_{\mu\nu}{\Box} R^2 $$
$$+ 9 ( - \frac{1}{2} g_{\mu\nu}{\Box} R^2 + 2 R_{\mu\nu}{\Box} R  + R^2_{; \mu\nu} -{\frac{1}{2}}g_{\mu\nu}R^3 + 3R^2 R_{\mu\nu})\} = 0 , \eqno(2.2)$$
where 
$${\Box} = \frac{1}{\sqrt{- g}}\frac{\partial}{\partial x^{\mu}}
\Big[
\sqrt{- g} g^{\mu\nu} \frac{\partial}{\partial x^{\nu}}\Big].
  \eqno(2.3)$$

General anastaz for space-time is given as

$$ dS^2 = g_{\mu\nu} d x^{\mu} d x^{\nu}  .  \eqno(2.4)$$

Trace of field equations (2.3) yields
$$\Big[ -
{\frac{1}{32{\pi}G }} + 2{\lambda}{\alpha} T^2\Big]R + {\alpha} ( 8 {\Box} R +
\frac{1}{2} R^2 ) + 8{\eta}^2 {\lambda}{\alpha} R^3 = 0.  \eqno(2.5)$$

$R$, being a linear combination of second-order derivatives  and square of
first-order derivatives of $g_{\mu\nu}$, has mass dimension 2, whereas the
same for scalar fields in known theories is 1. So, eq.(2.5) is  multiplied by
$\eta$ (having length dimension)and $\eta R$ is recognized $\tilde R$. Thus $\tilde R$, having mass
dimension 1, represents a scalar particle. This scalar particle is called
``riccion''. In the case of scalaron, $\eta$ is dimensionless and its
magnitude is 1 [6]. So, it is obtained that

$${\Box}{ \tilde R} + \frac{1}{16}R { \tilde R} + \frac{1}{8}\Big[ -
{\frac{1}{32{\pi}G {\alpha} }} + 2{\lambda} T^2\Big]{ \tilde R} + {\lambda}{\eta}^2 { \tilde R}^3 = 0    \eqno(2.6)$$

If ${\tilde R}$ is a basic
physical field, there should be an action $S_{\tilde R}$ yielding eq.(2.6) for invariance
of $S_{\tilde R}$, under transformations ${\tilde R} \to {\tilde R} +
\delta {\tilde R}$.

In what follows, $S_{\tilde R}$ is obtained. If such an action exists, one can write
$$\delta S_{\tilde R} = - \int {d^4 x} \sqrt{- g_4} \delta{\tilde R}[({\Box} + \frac{1}{16}R { \tilde R} + \frac{1}{8}\Big[ -
{\frac{1}{32{\pi}G {\alpha} }} + 2{\lambda} T^2\Big]{ \tilde R} + {\lambda}{\eta}^2 { \tilde R}^3 ] \eqno(2.7a)$$
which yields eq.(2.6) if $\delta S_{\tilde R} = 0$ under transformations
${\tilde R} \to {\tilde R} + \delta {\tilde R}.$ 

Eq.(2.7a) is re-written as
$$\delta S_{\tilde R}  =   \int {d^4 x} \sqrt{- g} \delta\Big[\frac{1}{2}\partial^{\mu}{\tilde R} \partial_{\mu}(\delta{\tilde R}) - \frac{1}{48 \eta}R { \tilde R}^2 - V^T({\tilde R}) \Big], \eqno(2.7b)$$
where 
$$ V^T({\tilde R}) = - {\frac{1}{2}}m^2({\tilde R}^2 + {\frac{1}{12}}T^2) +
{\frac{1}{4}}{\lambda}{\tilde R}^4 + {\frac{1}{8}}{\lambda}T^2{\tilde R}^2 -
{\frac{{\pi}^2}{90}}T^4  \eqno(2.7c)$$
with

$$ m^2 = \frac{1}{256 \pi G \alpha} . \eqno(2.7d)$$

$R, {\tilde R}$ and ${d^4 x} \sqrt{- g}$ are invariant under co-ordinate transformations. So, $R(x) = R(X), {\tilde R}(x) = {\tilde R}(X)$ and ${d^4 x} \sqrt{- g} = d^4 X,$ where $X^i (i=0,1,2,3)$ are local and  $x^i (i=0,1,2,3)$ are global coordinates. Moreover,
$${\Box} = g^{ij} \frac{\partial^2}{\partial x^i \partial x^j} + \frac{1}{2}  g^{mn} \frac{\partial g_{mn}}{\partial x^i} g^{ij}\frac{\partial}{\partial x^j} + \frac{\partial g^{ij}}{\partial x^i}\frac{\partial }{\partial x^j} = \frac{\partial^2}{\partial X^i \partial X^j} $$
in a locally inertial co-ordinate system, where $g_{ij} = \eta_{ij} $ (components of Minkowskian metric) and $g^{ij}_{,i} = 0$ (comma (,) stands for partial derivative). Thus, in a locally inertial co-ordinate system,
\begin{eqnarray*}
\delta S_{\tilde R} & = & \int {d^4X} \delta 
\Big[\frac{1}{2}\partial^{\mu}{\tilde R} \partial_{\mu}(\delta{\tilde R}) - \frac{1}{48 \eta}R { \tilde R}^2 - V^T({\tilde R}) \Big] \\ & = & \delta \int {d^4X} \Big[\frac{1}{2}\partial^{\mu}{\tilde R} \partial_{\mu}(\delta{\tilde R}) - \frac{1}{48 \eta}R { \tilde R}^2 - V^T({\tilde R}) \Big].
\end{eqnarray*}
$$ \eqno(2.8)$$

Employing principles of covariance and equivalence in  eq.(2.8)
$$\delta S_{\tilde R} = \delta \int {d^4 x} \Big\{\sqrt{- g}
\Big[\frac{1}{2}\partial^{\mu}{\tilde R} \partial_{\mu}(\delta{\tilde R}) - \frac{1}{48 \eta}R { \tilde R}^2 - V^T({\tilde R}) \Big] $$

yielding  the action for riccion as

$$ S_{\tilde R} =  \int {d^4 x} \Big\{\sqrt{- g}
\Big[\frac{1}{2}\partial^{\mu}{\tilde R} \partial_{\mu}(\delta{\tilde R}) - \frac{1}{48 }R { \tilde R}^2 - V^T({\tilde R}) \Big]. \eqno(2.9 ) $$

 It is important to mention here that  ${\tilde R}$ is different from other scalar fields due to dependence of (mass)$^2$ for ${\tilde R}$
on the gravitational constant and the
coupling constant $\alpha$ which is given by the eq.(2.1). Moreover, it emerges from geometry of the space-time.

Further, action for a scalar $\phi$ and a Dirac spinor $\psi$ are taken as
\begin{eqnarray*}
 S_{(m)} &=&   \int {d^4 x} \Big\{\sqrt{- g} [{1 \over2}
g^{\mu\nu}{\partial_{\mu}\Phi}{\partial_{\nu}\Phi^*}  -
\frac{1}{2}\Lambda {\tilde R}^2 \Phi\Phi^* \Big] \\&&+ {1 \over 2} \int {d^4x} \sqrt{ - g} [{\bar \psi} i \gamma^{\mu}
D_{\mu} \psi - \sigma {\tilde R}{\bar \psi} \psi + c.c.]  ,
\end{eqnarray*}
$$ \eqno(2.10)$$
where ${\Lambda}$ and $\sigma$ are  dimensionless coupling constants.Here ${\bar \psi} =
 \psi^{\dagger} {\tilde \gamma}^0, $   $D_{\mu}$ are covariant derivatives. $\gamma^{\mu}$ are curved space Dirac matrices and ${\tilde \gamma}^{\mu}$ being flat space Dirac matrices.  
  
So, the total action is
\begin{eqnarray*}
S &=& S_{\tilde R} + S_m \\ & =&   \int {d^4 x} \Big\{\sqrt{- g}
\Big[\frac{1}{2}\partial^{\mu}{\tilde R} \partial_{\mu}(\delta{\tilde R}) -
\frac{1}{48 }R { \tilde R}^2 - V^T({\tilde R}) \Big]\\&&  +  \int {d^4 x} \Big\{\sqrt{- g} [{1 \over2}
g^{\mu\nu}{\partial_{\mu}\Phi}{\partial_{\nu}\Phi^*}  -
\frac{1}{2}\Lambda {\tilde R}^2 \Phi\Phi^* \Big] \\&&+ {1 \over 2} \int {d^4x} \sqrt{ - g} [{\bar \psi} i \gamma^{\mu}
D_{\mu} \psi - \sigma {\tilde R}{\bar \psi} \psi + c.c.]  ,
\end{eqnarray*}
$$ \eqno(2.11)$$

\vspace{0.5cm}

\noindent \underline{\bf (b) Models of the early universe inspired by dual
  role }

\underline{\bf  of the Ricci scalar }

\bigskip

Here models of the early universe are obtained in vacuum states of riccion,
given by Higgs like potential (2.7c) [11, 17, 18] employing the condition

$$\frac{{\partial}V^T}{{\partial}{\tilde R}} = 0,$$
or $$-m^2{\tilde R} + {\lambda}{\tilde R}^3 +
{\frac{1}{4}}{\lambda}T^2{\tilde R} = 0 , \eqno(2.12)$$ 
This equation yields turning points of $V^T({\tilde R})$ as$$\tilde R = 0  \eqno(2.13a)$$and
$$\tilde R = \pm {\frac{1}{2}}{\sqrt(T^2_c - T^2)},  \eqno(2.13b)$$ where $$
T_c = {\frac{2m}{\sqrt{\lambda}}}  \eqno(2.13c)$$ 

These states  can be written in terms of vacuum
expectation value of $\tilde R$ as $<{\tilde R}> = 0 $ and $<{\tilde R}> =
(1/2){\sqrt{(T^2_c - T^2)}}.$ It shows that as long as $T\ge T_c$ , the
field $\tilde R$ remains confined to the state $<{\tilde R}> = 0.$ But
when temperature goes down such that $T \ll T_c$ , $\tilde R$ tunnels through
the temperature barrier $ T = T_c$ and settles in the state $<{\tilde R}>
= (1/2)T_c$ when $T\ll T_c.$  The state $<{\tilde R}> = 0 $ is called the
false vacuum and $<{\tilde R}>= (1/2)T_c$ corresponds to the true vacuum state
. Moreover , symmetry is intact in the state  $<{\tilde R}> = 0, $  but it is
broken spontaneously at  $<{\tilde R}>
= (1/2)T_c $ and energy is released as latent heat in the form
of radiation with density given as 
$$ V^T (<{\tilde R}> = 0) - V^T (<{\tilde R}> = {1 \over 2} T_c) =
\frac{m^4}{4 \lambda}. \eqno(2.14)$$

The line - element for the cosmological model of the early universe is
taken as $$ dS^2 = dt^2 - a^2(t)f^2(r)\Big[ dr^2 + r^2 (d{\theta}^2 +
sin^2{\theta}d{\psi}^2)\Big ] , \eqno(2.15a)$$ where $t$ is the cosmic time ,$0
\le r \le \infty,$
$0 \le {\theta} \le \pi$ and   $0 \le {\psi} \le 2\pi.$

Looking at  the line - element , given by (2.15a) , one may think that the
function 
$f$ can be absorbed by redefining $r$ as new coordinates ${\tilde r}$ such that 
$$f^2(r)\Big[  dr^2 + r^2 (d{\theta}^2 +
sin^2{\theta}d{\psi}^2)\Big] =  d{\tilde r}^2 + {\tilde r}^2 (d{\Theta}^2 +
sin^2{\Theta}d{\Psi}^2)\eqno(2.15b) .$$ It is important
to mention here that redefinition of coordinates are not valid at points
where $f = 0\quad{\rm or}\quad \infty ,$ as such points lead to spatial
singularity (singularity at points on $t = constant$ hypersurface).
So,the absorption of $f$ in $r$ coordinates is possible only
when $ f\ne 
0$ and $f<\infty$ at all points of $t = constant$ hypersurface. It means
that only after getting explicit definition of $f$ , it is possible to decide about $f$. This aspect will be looked into below.

Using geometrical aspect of the Ricci scalar , one can compute $\tilde R$
in the background geometry of the line - element , given by eq.(2.15) , as
$$ \tilde R = {\eta}\Big[6\Big\{{\frac{\ddot a}{a}} + \Big({\frac{\dot
a}{a}}\Big)^2\Big\} - {\frac{1}{a^2 r^2}}
\Big[{\frac{2}{ f^4}}\Big\{{\frac{{\partial}^2f^2}{{\partial}r^2}} +
{\frac{2}{r}}{\frac{{\partial}f^2}{{\partial r}}} \Big\} +
{\frac{3}{2 f^6}}\Big({\frac{{\partial}f^2}{{\partial}r}}\Big)^2 \Big]\Big] ,
\eqno(2.16) $$ 
which is 
linear sum of the homogeneous part , given as $6{\eta}\Big\{{\frac{\ddot
a}{a}} + 
\Big({\frac{\dot a}{a}}\Big)^2\Big\}$ as well as the inhomogeneous part
containing 
derivatives of $f^2.$ $<\tilde R>$ , being vacuum expectation value of
$\tilde R$ , is homogeneous . So , in the vacuum states
$<{\tilde R}>= 0$ and $<{\tilde R}> = \pm(1/2)T_c$,
$${\frac{1}{a^2 r^2}}
\Big[{\frac{2}{ f^4}}\Big\{{\frac{{\partial}^2f^2}{{\partial}r^2}} +
{\frac{2}{r}}{\frac{{\partial}f^2}{{\partial r}}} \Big\} +
{\frac{3}{2 f^6}}\Big({\frac{{\partial}f^2}{{\partial}r}}\Big)^2 \Big] = 0 ,
\eqno(2.17) $$ 

\smallskip

\noindent \underline{\bf  First  stage of the early universe }

\smallskip

Coming to the physical aspect of $\tilde R$, one finds that at $ T\ge T_c$,
$\tilde R$ stays in the false vacuum state  $\tilde R = <\tilde R> = 0.$
So , from eq.(2.16) , one obtains the differential equation
$${\frac{\ddot a}{a}} + \Big({\frac{\dot a}{a}}\Big)^2 = 0  \eqno(2.18)$$
which yields the solution [17]
$$ a^2 = a^2_c + {\frac{\mid t \mid}{t_{pl}}} , \eqno(2.19)$$
where $t_{pl}$ is the Planck time.
At this point, a question arises whether $ a_c = 0$ or it is non - zero .
In what follows, this question is answered  .

Components of energy - momentum tensor for matter field $\tilde R$ can be
defined as [11, 14-15] $$T_{\mu\nu} = 2{\frac{1}{\sqrt{-g}}}{\frac{{\delta}S_{\tilde
R}}{{\delta}g^{\mu\nu}}}$$ which is obtained as
$$T_{\mu\nu} = {\partial}_{\mu}{\tilde R}{\partial}_{\nu}{\tilde R} -
g_{\mu\nu}[{\frac{1}{2}}{\partial}^{\gamma}{\tilde
R}{\partial}_{\gamma}{\tilde R} - V^T({\tilde R})] -
2{\eta}R_{\mu\nu}{\Box}{\tilde R} + 2{\eta}m^2{\tilde R}R_{\mu\nu}$$ $$
-2{\eta}{\lambda}{\tilde R}^3 R_{\mu\nu} -
{\frac{1}{2}}{\eta}{\lambda}T^2{\tilde R}R_{\mu\nu} + 4{\eta}m^2{\tilde
R}_{;\mu\nu} - 4{\eta}m^2g_{\mu\nu}{\Box}{\tilde R}$$ $$ -
4{\eta}{\lambda}{\tilde R}^3_{;\mu\nu} +
4{\eta}{\lambda}g_{\mu\nu}{\Box}{\tilde R} - {\lambda}T^2{\tilde
R}_{;\mu\nu} + {\lambda}T^2g_{\mu\nu}{\Box}{\tilde R}.  \eqno(2.20)$$

In the false vacuum state  $\tilde R = <\tilde R> = 0 ,$ one obtains  [11,16, 17]$$
T_{\mu\nu}U^{\mu}U^{\nu} < 0 , \eqno(2.21)$$ where $
T_{\mu\nu}$ are given by eq.(2.20) and $U^{\mu}$ are
velocities normalized as $$U^{\mu}U_{\mu} = +1 .$$ 
The result, given by the
inequality (2.21),shows that the energy condition is voilated in the false
vacuum state when $T\ge T_c$ [11,15,17]. It means that the universe will bounce without
any encounter 
with temporal singularity (singularity in time). As a
result , $$ a_c \ne 0 . \eqno(2.22)$$ 

The partial differential equation
(2.17) yields an exact solution 
$$ f^2 = \Big[ 1 - \frac{r_0}{r}\Big]^{4/7} . \eqno(2.23) $$ 

Thus, in the false vacuum state, model of the early universe is obtained
as given by eq.(2.15a) with $a^2(t)$  and $f^2(r)$ defined by eqs.(2.19) and
(2.23) respectively.

The expansion of the cosmological model, derived above, is adiabatic. So, entropy will remain conserved , which implies that the
average uniform temperature will fall as $$ T { \propto }\Big[ a_c^2 +
({\mid t \mid}/t_pl)\Big]^{-1/2} . \eqno(2.24) $$ 

\smallskip

\noindent \underline{\bf  Second stage of the early universe }

\smallskip

When $ T \ll T_c$ 
$${\tilde R} = < 
{\tilde R} > = \pm (1/2) T_c . $$ 
In this state also , one
obtains two differential equations $$ 6{\eta}\Big\{{\frac{\ddot a}{a}} +
\Big({\frac{\dot a}{a}}\Big)^2\Big\} = \pm {\frac{1}{2}} T_c
\eqno(2.25)$$ and the other
differential equation is the equation given by eq.(2.17). Eq.(2.25) yields two
cases. 
The first case is   $$ 6{\eta}\Big\{{\frac{\ddot a}{a}} +
\Big({\frac{\dot a}{a}}\Big)^2\Big\} = {\frac{1}{2}} T_c ,\eqno(2.26) $$
yielding the solution 
\begin{eqnarray*}
a^2(t) &=& {\sqrt{\frac{24}{{\eta}T_c}}} sinh\Big[(t -
t_{1e}){\sqrt{\frac{T_c}{6{\eta}}}} + D\Big] \\ & =& a^2_{1e} [sinh
D]^{-1} sinh\Big[(t - t_{1e}){\sqrt{\frac{T_c}{6{\eta}}}} + D\Big] \\&&
\simeq a^2_{1e}  exp\Big[(t - t_{1e}){\sqrt{\frac{T_c}{6{\eta}}}} \Big]
\end{eqnarray*}
$$ \eqno(2.27a)$$ 
with $a_{1e} = a ( t = t_{1e})$ and
$$ D  =  sinh^{-1} \Big[a^2_{1e} {\sqrt{\frac {{\eta}T_c}{24}}} \Big] .$$
Here $t_{1e}$ is the time at the end of first stage of the universe and
$a_{1e}$ is the corresponding scale factor, given as.

$$ a(t) \simeq a_{1e} exp\Big[(t - t_{1e}){\sqrt{\frac{T_c}{24{\eta}}}} \Big] \eqno(2.27b)$$

The second  case is 
 $$ 6{\eta}\Big\{{\frac{\ddot a}{a}} + \Big({\frac{\dot
a}{a}}\Big)^2\Big\} = - {\frac{1}{2}} T_c  \eqno(2.28)$$ 
which 
integrates to  $$ a^2(t) = a^2_{10} sin\Big[(t -
t_0){\sqrt{\frac{T_c}{6{\eta}}}} + {\frac{\pi}{2}}\Big] . \eqno(2.29)$$ In
true vacuum states also ,$ f$ is given by eq.(2.23). 

But $a(t),$ given by
eqs.(2.27), shows that the cosmological model will expand more rapidly in
the state  $ {\tilde R} = <{\tilde R} > = (1/2) T_c$ compared to the state
$ {\tilde R} = <{\tilde R} > = 0$. The expansion , in the state   $
{\tilde R} = <{\tilde R} > = - (1/2) T_c$, is given by eq.(2.29), which
shows that $ - a^2_{10} \leq a^2 \leq   a^2_{10}$ leading to imaginary
$a(t)$ also, which is unphysical. So the state $<{\tilde R} > = - (1/2)
T_c$ will not be considered any more.

 Thus one finds that vacuum states
${\tilde R} = <{\tilde R} > = 0$ and   $
{\tilde R} = <{\tilde R} > = (1/2) T_c$ lead to  cosmological models  $$
dS^2 = dt^2 - \Big[ a^2_c + {\frac{\mid t \mid}{t_{Pl}}}\Big] \Big[ 1 -
\frac{r_0}{r}\Big]^{4/7} \Big[ dr^2 + r^2 (d{\theta}^2 +
sin^2{\theta}d{\psi}^2)\Big ] \eqno(2.30)$$ at temperature $ T
\ge T_c$ and   $$ dS^2 = dt^2 - a^2_{1e}
exp\Big[(t - t_{1e}){\sqrt{\frac{T_c}{6{\eta}}}} \Big] \Big[ 1 -
\frac{r_0}{r}\Big]^{4/7} \Big[ dr^2 + r^2 (d{\theta}^2 + 
sin^2{\theta}d{\psi}^2)\Big ] \eqno(2.31)$$ 
when $ T\ll T_c.$

Divergence of scalar polynomials of the curvature tensors  at points $r =
0$ and $r = r_0.$ shows that  
models ( given by eqs.(2.30) and (2.31)) are singular at these points [21-22]. Moreover, these models are inhomogeneous at small scales i.e. when $r$ is comparable to $r_0$ and homogeneous when $r>> r_0.$

The model (2.31) exhibits accelerated expansion of the universe which is
consistent with consequences of recent experiments Ia Supernova [24] and WMAP [20].

It is interesting to note that models, given by eqs.(2.30) and (2.31), reduce
to homogeneous Robertson-Walker type models as $ r \to \infty.$ As far as
spatial singularity is concerned,  at a
particular instant of time , these models can be compared with
Scwarzschild space-time [19, 21]
$$ dS^2 = \Big[ 1 - \frac{2GM}{r} \Big]dt^2 -  \Big[ 1 - \frac{2GM}{r}
\Big]^{-1} dr^2 -  r^2 \Big[d{\theta}^2 + sin^2{\theta}d{\psi}^2)\Big]
\eqno(2.34)$$ 

It is remarked that the Scwarzschild space-time given by eq.(2.34)is
singular at  $r = 0$ and $r = 2GM$, but only $r = 0$ is the real physical
singularity and the other point is a co- ordinate singularity. In
models , given by eqs.(2.30) and (2.31), both points $r = 0$ and $r = r_0$
exhibit 
real  singularities as mentioned above. Physically  $ r_0$ signifies radius of particles or  compact
objects dominating the  expanding universe. Further investigations are carried out for $r > r_0.$

\vspace{0.5cm}

\centerline{\bf 3. Creation of spinless particles}

\bigskip

Invariance of $S$, given by eq.(2.11) under transformation $\Phi \to \Phi + \delta \Phi$ leads
to the Klein - Gordon equation
$$[{\Box} + M^2_b ] \Phi = 0 ,   \eqno(3.1)$$ 
where 
$$ M^2_b = \Lambda <{\tilde R}>^2 .
\eqno(3.2)$$ 

In the background geometry of the line - element (2.15 a) with eq.(2.23), the
equation (3.1) is re-written as
$$\frac{\partial^2 \Phi}{\partial t^2}  + \frac{3{\dot a}}{a}
\frac{\partial \Phi}{\partial t} - {1 \over a^2}\Big[ 1 - \frac{r_0}{r}
\Big]^{-4/7} \frac{\partial^2 \Phi}{\partial r^2}   - {1 \over r^2 a^2}\Big[ 1
- \frac{r_0}{r} \Big]^{-6/7} \frac{\partial}{\partial r}\Big[ r^2 \Big( 1
- \frac{r_0}{r}\Big)^{2/7}  \Big] \frac{\partial \Phi}{\partial r}$$
$$- {1 \over a^2 r^2}\Big[ 1 - \frac{r_0}{r}
\Big]^{-4/7} \Big[\frac{1}{sin \theta} \frac{\partial }{\partial \theta}
\Big({sin \theta} \frac{\partial }{\partial \theta} \Big) + \frac{1}{sin^2
\theta} \frac{\partial^2 }{\partial \psi^2} \Big] \Phi + M^2_b \Phi = 0,
\eqno(3.3)$$ 
where dot (.) over the variable denotes the derivative with respect to time.

The solution for this equation is taken as
$$\Phi = \Big[ (2 \pi )^{1/2} r \Big( 1 - \frac{r_0}{r} \Big)^{3/7} \Big]^{-1}
{\sum_{k,l,m}}  \Big[ A_{k l m} \Psi_{k l m} (t) Y_{l m}(
\theta, \psi) e^{ i k.r} + c.c. \Big],  \eqno(3.4)$$
where $r> r_0, m = - l \cdots +l, l = 1,2,3, \cdots and -\infty < k < \infty.$
$Y_{l m}(\theta, \psi)$ satisfies the equation
$$\Big[\frac{1}{sin \theta} \frac{\partial }{\partial \theta}
\Big({sin \theta} \frac{\partial }{\partial \theta} \Big) + \frac{1}{sin^2
\theta} \frac{\partial^2 }{\partial \psi^2} \Big] Y_{l m} = - l (l + 1)
Y_{l m} \eqno(3.5a)$$
with normalization as
$$ \int sin \theta { d \theta}{d \psi} Y_{l m} Y_{l^{\prime} m^{\prime}} =
\delta_{l l^{\prime}} \delta_{m m^{\prime}}. \eqno(3.5b)$$

Connecting eq.(3.3) with eq.(3.4) and using eq.(3.5), it is obtained that
$$ e^{i k.r} \Big[{\ddot \Psi_{k l m}} + \frac{3 {\dot a}}{a} {\dot
\Psi_{k l m}} + \Big(\Big\{\frac{k^2}{a^2} + \frac{ l (l + 1)}{a^2 r^2} +
 \frac{ 6 r_0 ( 2 r - r_0 )}{7 a^2 r^2 ( r - r_0 )^2}\Big\} \Big[ 1
- \frac{r_0}{r} \Big]^{-4/7}$$ $$ + M^2_b \Big) \Psi_{k l m} \Big]  = 0
.  \eqno(3.6)$$ 

Using the convolution theorem [25] and integrating over $r$, eq.(3.6) looks like
$$ {\int^{\infty}_{r_0 + \epsilon}}e^{i k.r} {dr} \Big[{\ddot \Psi_{k l m}} +
\frac{3 {\dot a}}{a} {\dot 
\Psi_{k l m}} + \Big\{\frac{k^2}{a^2 \eta } {\int^{\infty}_{r_0 + \epsilon}}
e^{ - i k.y} \Big[ 1 - \frac{r_0}{y}
\Big]^{- 4/7}{dy} $$ $$+ \frac{ l (l + 1)}{a^2 \eta} {\int^{\infty}_{r_0 +
\epsilon}} \frac{e^{ - i k.y}}{y^2} \Big[ 1 - \frac{r_0}{y}
\Big]^{-4/7} {dy} + \frac{6r_0}{7 a^2 \eta}{\int^{\infty}_{r_0 +
\epsilon}} \frac{( 2 y - r_0 )}{ y^2 ( y - r_0 )^2} \Big[ 1 
- \frac{r_0}{y} \Big]^{-4/7} {d y}$$ $$+ M^2_b \Big\} \Psi_{k l m} \Big]  = 0
.  \eqno(3.7)$$ 
Here $\epsilon$ is an extremely small positive real number.

Details of the evaluation of integrals with respect to $y$ are given in
the Appendix B. Using these results, in eq.(3.7), one obtains
$${\ddot \Psi_{k l m}} + \frac{3 {\dot a}}{a} {\dot \Psi_{k l m}} + \Big[
\frac{X_{k l m}}{a^2} + M^2_b \Big] \Psi_{k l m} = 0 , \eqno(3.8a)$$
where
$$X_{k l m} = ( 1/a^2 \eta ) \epsilon^{3/7} (r_0 + \epsilon)^{4/7}
cos \{ k (r_0 + \epsilon)\} 
\Big[  k^2 + \frac{ l ( l + 1)}{ ( r_0 + \epsilon )^2}$$  $$ -
\frac{12}{7} r_0 \epsilon^{-2} (r_0 + \epsilon)^{-1} + \frac{6}{ 7} r_0^2
\epsilon^{-2} ( r_0 + \epsilon )^{-2} \Big] . \eqno(3.8b)$$

\bigskip

\noindent \underline{Case 1: The case of state $ < {\tilde R}> = 0$}

 \bigskip
 
Using $a(t)$ from the line-element (2.30) for this state, the equation (3.8) is
written as

$${\ddot \Psi_{k l m}} + \frac{3 }{ 2 ( t + a_{10}^2 t_P)} {\dot \Psi_{k l
m}} + \frac{X_{k l m} t_P}{ (t + t_P a_{10}^2)}  \Psi_{k l m} = 0 ,
\eqno(3.9)$$ 
which yields the solution for $t>0$ as
$$ \Psi_{k l m} = \tau^{ - 1/2} \Big[ A J_{-1/2} (\tau) + B Y_{-1/2}
(\tau) \Big] , \eqno(3.10a)$$
where $A$ and $B$ are integration constants. Here $\tau$ is defined as
$$ \tau =  \sqrt{t_P X_{k l m}  ( t + t_P a_{10}^2 )}  \eqno(3.10b)$$
showing that $\tau \to \infty$ when $t \to \infty.$

For large $\tau$,
$$J_{-1/2} (\tau) \simeq \frac{ cos (\tau)}{\sqrt{\pi\tau/2}} \quad{\rm
and} \quad Y_{-1/2} (\tau) \simeq \frac{ sin (\tau)}{\sqrt{\pi\tau/2}}.$$
So, when $\tau$ is large,
\begin{eqnarray*}
\Psi_{k l m} &=&  [ \pi \tau^2/2]^{-1/2} [ A cos \tau  +  B sin \tau] \\ &
= & \Big[ \frac{\tau_1^3}{2 t_P X_{k l m}}\Big]^{1/2} \tau^{-1} \Big[(1 + i)
e^{-i \tau} + (1 - i) e^{i \tau} \Big]
\end{eqnarray*}
$$ \eqno(3.11)$$   
using the normalization condition
$$ (\Phi_{k l m} , \Phi_{k l m}) = 1  = - (\Phi^*_{k 
l m} , \Phi^*_{k l m}), \eqno(3.12a)$$ 
where 
$$\Phi_{k l m} = \Big[( 2 \pi \eta^{-1} W)^{1/2} r \Big( 1  - \frac{r_0}{r} \Big)^{1/7}
\Big]^{-1} \Psi_{k l m} (t) e^{ i k.r} Y_{lm} \eqno(3.12b)$$  
and the scalar product 
is defined as
$$ (\Phi_{k l m} , \Phi_{k l m}) = -
i {\int_{r_0 + \epsilon}^{\infty}}{\int_0^{\pi}}
{\int_0^{2\pi}}\sqrt{ - g_{\Sigma}}  {d{\Sigma}} [ 
\Phi_{k l m} (\partial_t \Phi^*_{k^{\prime}l^{\prime} m^{\prime}})$$  $$ -
(\partial_t \Phi_{k l m}) \Phi^*_{k^{\prime}l^{\prime} m^{\prime}}] |Y_{l m}(
\theta, \psi)|^2 , 
\eqno(3.12c)$$ 
where $\Sigma$ is the $t = t_1$ hypersurface, $\sqrt{ - g_{\Sigma}} = r^2
\Big[1 - \frac{r_0}{r} \Big]^{6/7}$  and $ d\Sigma = sin \theta dr
d\theta d\psi.$ 

Thus
$$\Psi^{\rm out}_{k l m} = \Big[ \frac{\tau_1^3}{2 t_P X_{k l
m}}\Big]^{1/2} \tau^{-1} (1 + i) e^{-i \tau}.  \eqno(3.13)$$  

For $t<0,$ eq.(3.9) yields the solution
$$ \Psi_{k l m} = {\tilde \tau}^{ - 1/2} \Big[ A^{\prime} J_{-1/2}
({\tilde \tau}) 
+ B^{\prime} Y_{-1/2} ({\tilde \tau}) \Big] , \eqno(3.14a)$$
where $A^{\prime}$ and $B^{\prime}$ are integration constants. Here $\tilde \tau$
is defined as $$ {\tilde \tau} = -  \sqrt{t_P X_{k l m} (- t + t_P a_0^2
)}  \eqno(3.14b)$$ 
showing that $\tilde \tau \to - \infty$ when $t \to - \infty.$

Using above approximations for Bessel's functions and normalization
prescription, it is obtained that for 
$t \to - \infty$
$$\Psi^{\rm in}_{k l m} =  \Big[ \frac{{\tilde \tau}_1^3}{2 t_P X_{k l
m}}\Big]^{1/2} {\tilde \tau}^{-1} (1 + i) e^{+i {\tilde \tau}}.
\eqno(3.15)$$   

Since $\Phi^{\rm out}_{k l m}$ and $\Phi^{\rm in}_{k l m}$ both belong to
the same Hilbert space, so one write
\begin{eqnarray*}
\Phi &=& {\sum_{k,l,m}}  \Big[ A^{\rm in}_{k l m} \Phi^{\rm in}_{k
l m}  Y_{l m}( \theta, \psi)  +   A^{{\rm in}\dagger}_{k l m} \Phi^{*{\rm
in}}_{k l m}  Y_{l m}( \theta, \psi)\Big] \\ &=&  
{\sum_{k,l,m}}  \Big[ A^{\rm out}_{k l m} \Phi^{\rm out}_{k l m}
 Y_{l m}( \theta, \psi)  +   A^{{\rm out}\dagger}_{k l m} \Phi^{*{\rm
out}}_{k l m}  Y_{l m}( \theta, \psi)\Big] .
\end{eqnarray*}

As a result, one obtains
$$\Phi^{\rm out}_{k l m} = \alpha_{k l m} \Phi^{\rm in}_{k l m}  +
\beta_{k l m} \Phi^{*{\rm in}}_{k l m} , \eqno(3.16a)$$  
where $\alpha_{k l m}$ and $\beta_{k l m}$ are Bogoliubov coefficients
satisfying the condition [ 18, 26-27]
$$|\alpha_{k l m}|^2  - |\beta_{k l m}|^2  = 1. \eqno(3.16c)$$

The in- and out- vacuum states are defined as
$$A^{\rm in}_{k l m}|{\rm in}> = 0 = A^{\rm out}_{k l m}|{\rm out}>
.\eqno(3.17a,b)$$ 
Moreover,
$$A^{\rm out}_{k l m} = \alpha_{k l m} A^{\rm in}_{k l m}  +
\beta_{k l m} A^{*{\rm in}}_{k l m} , \eqno(3.17c)$$

Connecting eqs.(3.11) - (3.17), it is obtained that
\begin{eqnarray*}
\alpha_{k l m} & =& {1 \over 2} e^{ - i ( \tau + {\tilde \tau})}\Big[
\Big( \frac{ t_1 + a_{10}^2 t_P}{ - t_1 + a_{10}^2 t_P}\Big)^{1/4} + \Big(
\frac{ - t_1 + a_{10}^2 t_P}{  t_1 + a_{10}^2 t_P}\Big)^{1/4}\Big] \\
\beta_{k l m} & =& {1 \over 2} e^{ - i ( \tau + {\tilde \tau})}\Big[ 
\Big( \frac{ t_1 + a_{10}^2 t_P}{ - t_1 + a_{10}^2 t_P}\Big)^{1/4} - \Big(
\frac{ - t_1 + a_{10}^2 t_P}{  t_1 + a_{10}^2 t_P}\Big)^{1/4}\Big]  .
\end{eqnarray*}
$$\eqno(3.18a,b)$$ 
These results yield
$$ \beta_{k l m} \simeq i\frac{a_{10}^2 t_P}{ t_1} e^{-i(\tau_1 + {\tilde
\tau}_1)}   \eqno(3.19a)$$ 
implying
$$ |\beta_{k l m}|^2 \simeq \Big[\frac{a_{10}^2 t_P}{ t_1}\Big]^2 =
\Big[\frac{5 t_P}{ t_1}\Big]^2.  
\eqno(3.19b)$$

 Eq.(3.19b) shows that when $t_1$ is sufficiently larger than $5 t_P$, $
 |\beta_{k l m}|^2 = 0.$ It means that spinless particles will  be created in the state
$<{\tilde R}> = 0$ only when $t_1 \le 5 t_P$. For $t_1$ sufficiently larger
 than $5 t_P$, there will be no production of scalar particles, in this state.

Anamolous terms emerge when vacuum expectation value of the
energy-momentum tensor components $T_{\mu\nu}$ is evaluated for $\Phi$, if
dimension of the space-time is an even integer. These terms get cancelled
for the difference of vacuum 
expectation values $T_{\mu\nu}$ in  in- and out-states. Thus $T_{\mu\nu}$
for created particles is defined as [28]
$$<T_{\mu\nu}> =   <{\rm out}|
T_{\mu\nu}|{\rm out}> - <{\rm in}| T_{\mu\nu}|{\rm in}> .  \eqno(3.20a)$$

 $T_{\mu\nu}$ for $\phi$ are obtained from the  action (2.11) as
$$T_{\mu\nu} = {\partial_{\mu}\Phi^*}{\partial_{\nu}\Phi}  - 2 \eta
{\Lambda}[R_{\mu\nu} + \{ ({\tilde R}\Phi^* \Phi)_{; \mu\nu} - g_{\mu\nu}
{\Box}({\tilde R}\Phi^* \Phi) \} $$
$$  - {1 \over2}g_{\mu\nu}[{\partial^{\rho}\Phi^*}{\partial_{\rho}\Phi} - {\Lambda}{\tilde R}^2 \Phi^* \Phi ]  \eqno(3.20b)$$
with its trace
$$ T = - ( 1 -  12 \eta {\Lambda}{\tilde R}
){\partial^{\mu}\Phi^*}{\partial_{\mu}\Phi}  - 12 \eta \Lambda^2 <{\tilde R} >^3 \Phi^* \Phi  . \eqno(3.20c)$$

Now trace of $T_{\mu\nu}$ for created partcle-antiparticle pairs is obtained
as

$$<T> =<{\rm out}| T |{\rm out} > - <{\rm in}| T |{\rm in} > \eqno(3.20d)$$

Eqs.(3.12b), (3.13), (3.15) and (3.20 d) lead to trace of created particles in the state $<{\tilde R}> = 0$ as

$$ <T_1(b)> = 0 . \eqno(3.21)$$
 
\bigskip

\noindent \underline{Case 2: The case of state $ < {\tilde R}> =  {1 \over
2} T_c$}

 \bigskip

Using $a(t)$ from the line-element (2.31) for this state, the equation (3.8) is
written as

$${\ddot \Psi_{k l m}} + \frac{3 }{ 2} \sqrt{\frac{T_c}{6 \eta}} {\dot
\Psi_{k l m}} + \Big[ \frac{X_{k l m}}{a_0^2} e^{ - (t -
t_{1e})  \sqrt{T_c/6 \eta}}  + M^2_b \Big]  \Psi_{k l m} = 0 ,
\eqno(3.22)$$ 
where $M^2_b = {1 \over 4}\Lambda T_c^2$ using the definition of $M_b$ from eq.(3.2). The equation (3.22) yields the
solution for $t>0$ as
$$ \Psi_{k l m} = \Big[-\frac{\pi}{sin h 2 \pi \alpha}\Big]^{1/2}\Big[
\frac{6 \eta }{ T_c}\Big]^{1/4}
e^{ - \frac{3}{4} (t - t_{10}) \sqrt{ T_c/6 \eta}}\times$$ $$ J_{\pm 2i \alpha} \Big(
\gamma_{k l m} e^{ - {1 \over 2} (t - t_{10}) \sqrt{T_c/6 \eta}} \Big),
\eqno(3.23a)$$ 
where 
$$\alpha^2 = \frac{6 \eta}{T_c} \Big[{1 \over 4}\Lambda T_c^2 - \frac{3
T_c}{32 \eta} \Big] \eqno(3.23b)$$
and
$$\gamma^2_{k l m} = \frac{24 \eta X_{k l m}}{ a_0^2 T_c} .\eqno(3.23c)$$ 

For $t<0,$ the equation (3.22) yields the
solution  as
$$ \Psi_{k l m} =  \Big[-\frac{\pi \alpha}{\eta sin h (2 \pi \alpha)}\Big]^{1/2}\Big[
\frac{6 \eta }{ T_c}\Big]^{1/4} e^{\frac{3}{4} (t - t_{1e}) \sqrt{ T_c/6
\eta}}\times $$ 
$$J_{\mp 2i \alpha} \Big( \gamma_{k l m} e^{  {1 \over 2}(t - t_{1e}) \sqrt{T_c/6
\eta}} \Big). \eqno(3.24)$$

Using the approximation of
 $$J_n (x) \simeq  \frac{x^n}{ 2^n \Gamma (1 + n)}$$
for small $x$ in eqns.(3.23) and (3.24) ,
for $t \to \infty$
$$ \Psi^{\rm out}_{k l m} =  \Big[- \frac{\pi \alpha}{\eta sin h (2 \pi
\alpha})\Big]^{1/2}\Big[ \frac{6 \eta }{ T_c}\Big]^{1/4}
\Big(\frac{1}{\Gamma(1\pm 2 i \alpha)}\Big) 
\Big(\frac{\gamma_{k l m}}{2}\Big)^{\pm 2 i \alpha}\times$$ $$
e^{- (t-t_{1e}) (\frac{3}{4} \mp  i \alpha )\sqrt{ T_c/6 \eta}} , 
\eqno(3.25a)$$ 
and for $t \to -\infty$
$$ \Psi^{\rm in}_{k l m} =   \Big[-\frac{\pi \alpha}{\eta sinh 2 \pi
\alpha}\Big]^{1/2}\Big[  \frac{6 \eta }{ T_c}\Big]^{1/4}
 \Big(\frac{1}{\Gamma(1\mp 2 i \alpha)}\Big)
\Big(\frac{\gamma_{k l m}}{2}\Big)^{\mp 2 i \alpha} \times$$ $$ e^{(t -
t_{1e})(\frac{3}{4} \mp  i \alpha )\sqrt{ T_c/6 \eta} }, 
\eqno(3.25b)$$ 

Using $ \Psi^{\rm out}_{k l m} $ and $ \Psi^{\rm in}_{k l m}$, one obtains
$$\alpha_{k l m} = - i \Big(\frac{3}{2} \mp 2 i
\alpha  \Big)   \eqno(3.26a)$$  
and
$$\beta_{k l m} = - \frac{3i}{2}  \eqno(3.26b)$$ 
implying that $$|\beta_{k l m}|^2  = \frac{9}{4} , \eqno(3.26c)$$ 
which ensures creation of spinless particles.
Connecting eqs.(3.16c) and (3.26b)

$$|\alpha_{k l m}|^2 = \frac{13}{4} \eqno(3.26d)$$ 
Moreover, the absolute probability for no particle creation is given as
$$|<{\rm out}|{\rm in}>|^2 = {\prod_{k l m}} |\alpha_{k l m}|^{-2} =
\frac{4 }{9 + 16 \alpha^2}.  \eqno(3.27)$$ 

Eqs.(3.26 d) and (3.27) yield

$$ \alpha^2 = 1/4 .   \eqno(3.28)$$ 

Connecting eqs.(3.2), (3.23b) and (3.28), it is obtained that

$$ M_b^2 = \frac{13}{16} \Big( \frac{T_c}{6 \eta} \Big). \eqno(3.29)$$ 

Using the convolution theorem, eq.(3.12b) can be written as

\begin{eqnarray*}
\Phi_{klm}(t, r, \theta, \phi) &=& \Big[ (2 \pi)^{-1/2} e^{- ikr} \int_{r_0 +
  \epsilon}^{\infty} \Big\{ y \Big(1 - \frac{r_0}{y} \Big) \Big\}^{-1} e^{iky}
  {dy} \Big] \Psi_{klm} (t) Y_{lm} (\theta, \phi) \\ &=& (2 \pi)^{-1/2} e^{-
  ikr}\frac{\epsilon^{4/7}}{(r_0 +  \epsilon)^{11/7}} cos[k(r_0 +  \epsilon)] \Psi_{klm} (t) Y_{lm} (\theta, \phi).
\end{eqnarray*}
$$ \eqno(3.30)$$

Connecting eqs.(2.27 a), (3.20 d), (3.25) and (3.30) as well as taking average
over $\theta$ and $\phi$, trace of the energy-momentum tensor for created
particles, in the  state $ {\tilde R} = (1/2) T_c,$ is obtained at $t =
t_{2e}$ as

\begin{eqnarray*}
T_{2(b)} &=& \sum_{klm}(1/2 \pi) \Big(\frac{6 \eta}{T_c} \Big)^{1/2} sinh
[(3/4)\sqrt{\frac{T_c}{6 \eta}} (t_{2e} - t_{1e}) ] \\&&
\times\frac{\epsilon^{8/7}}{( r_0 + \epsilon )^{22/7}} cos^2 k(r_0 + \epsilon
) \Big[(13/16) \frac{T_c}{6  \eta} ( 6\eta \Lambda T_c - 1 ) - (3/2) \eta
\Lambda^2 T_c^3 \Big] \\ &=& (2/3 \pi) Big(\frac{6 \eta}{T_c} \Big)^{1/2} sinh
[(3/4)\sqrt{\frac{T_c}{6 \eta}} (t_{2e} - t_{1e}) ] \\&&
\times\frac{\epsilon^{8/7}}{( r_0 + \epsilon )^{22/7}} cos^2 k(r_0 + \epsilon
) \Big[(13/16) \frac{T_c}{6  \eta} ( 6\eta \Lambda T_c - 1 ) - (3/2) \eta
\Lambda^2 T_c^3 \Big]  . 
\end{eqnarray*}
$$ \eqno(3.31)$$

Here,  summation is done with the help of the Riemann
zeta function defined as $\zeta(s) = {\sum_{n=1}^{\infty}} n^{-s}.$  Moreover,
\begin{eqnarray*}
{\sum_{k l m}}cos^2 [k(r_0 + \epsilon)] &=& {\sum_{k l m}}[1 - k^2 (r_o + \epsilon)^2 +
\cdots]\\ & = & {\sum_{k l m}} 1 = {\sum_{k l }}(2 l + 1)\\ &=&{\sum_{k  }}[4 \zeta(-1) + 2\zeta(0)] = - 2(4/3)  \zeta(0) = 4/3
\end{eqnarray*}
as $\zeta (-2m) = 0$. Also $\zeta (0) = - {1 \over 2}$ and $\zeta (-1 ) = - {1 \over 12}$ obtained through the analytic continuation.

\vspace{0.5cm}

\centerline{\bf 4. Creation of spin-1/2 particles}

\bigskip

Using invariance of $S$, given by eq.(2.11) under transformation $\psi \to \psi + \delta \psi$, the Dirac equation is obtained as

$$ \Big( i \gamma^{\mu}D_{\mu}  -  M_f  \Big) \psi = 0,
\eqno(4.1a)$$ 
where
 
$$ M_f = \sigma {\tilde R}.\eqno(4.1b)$$

Here,    $D_{\mu} = \partial_{\mu} -
\Gamma_{\mu}$ with
+$$ \gamma^{\mu} = e^{\mu}_a {\tilde \gamma}^a,  \eqno(4.2a)$$
where $(a , \mu = 0,1,2,3)$ and $e^{\mu}_a$ are defined through 
$$e^{\mu}_a e^{\nu}_b g_{\mu\nu} = \eta_{ab}. \eqno(4.2b)$$
Here $\eta_{ab}$ are Minkowskian metric tensor components and $g_{\mu\nu}$
are metric tensor components in curved space-time. $c.c.$ stands for
complex conjugation.

Dirac matrices $\gamma^{\mu}$ in curved space-time satisfy the anti-commutation
rule [18]
$$ \{ \gamma^{\mu} , \gamma^{\nu} \} = 2 g^{\mu\nu}  \eqno(4.2c)$$
and Dirac matrices ${\tilde \gamma}^a$ in Minkowskian space-time 
satisfy the anti-commutation rule
$$ \{ {\tilde \gamma}^a , {\tilde \gamma}^b \} = 2 \eta^{ab}.
\eqno(4.2d)$$ 
$\Gamma_{\mu}$ are defined as
$$\Gamma_{\mu} = - {1 \over 4} \Big(\partial_{\mu}e^{\rho}_a +
\Gamma^{\rho}_{\sigma\mu} e^{\sigma}_a \Big) g_{\nu\rho} e^{\nu}_b{\tilde
\gamma}^b {\tilde \gamma}^a.  \eqno(4.2e)$$

In general, solution of the Dirac equation $\psi$ can be written as
$$ \psi = {\sum_{s=\pm1}}{\sum_k} \Big( b_{k,s} \psi_{I k,s} +
d^{\dagger}_{-k,-s} \psi_{II k,s} \Big)   , \eqno(4.3a)$$
$$ \psi^{\dagger} = {\sum_{s=\pm1}}{\sum_k} \Big(  {\bar \psi}_{I
k,s} \gamma^0 b_{k,s}^{\dagger}+
 {\bar \psi}_{II k,s} \gamma^0 d_{k,s} \Big)   , \eqno(4.3b)$$
where
$$\psi_{I k,s} =  \Big[ (2 \pi \eta)^{1/2} r \Big( 1 - \frac{r_0}{r}
\Big)^{1/7} \Big]^{-1} {\sum_{l,m}}  f_{k l m ,s} (t) Y_{l m}(
\theta, \phi) e^{ i k.r} u_s   \eqno(4.4a)$$
and
$$\psi_{II k,s} =  \Big[ (2 \pi  \eta)^{1/2} r \Big( 1 - \frac{r_0}{r}
\Big)^{1/7} \Big]^{-1} {\sum_{l,m}}  g_{k l m ,s} (t) Y_{l m}(
\theta, \phi) e^{-i k.r} {\hat u}_s .  \eqno(4.4b)$$
with $g_{k l m ,s} (t) = f_{- k l m ,s} (t)$.

In eqs.(4.4)
$$ u^{T}_1 = (1 0 0 0), \quad u^{T}_{-1} = (0 1 0 0)       $$
$$ {\hat u}^{T}_1 = (0 0 1 0), \quad {\hat u}^{T}_{-1} = (0 0 0 1) ,
\eqno(4.5a,b,c,d)      $$ 
where the index $(T)$ stands for transpose of the column matrices $u_{\pm
s}$ and ${\hat u}_{\pm s}$.
In eqs.(4.4), $ m = -l, \cdots,+l ; l = 1,2,3, \cdots$  and  $ - \infty <
k < \infty.$

Using eqs.(3.5) $\psi$ is normalized as
\begin{eqnarray*}
(\psi_{k,s}, \psi_{k^{\prime},s^{\prime}}) &=& {\int_{\Sigma}} {d^3x} {\bar
\psi}_{k,s} {\tilde \gamma}^0 \psi_{k^{\prime},s^{\prime}} \\ &=&
\delta_{kk^{\prime}} \delta_{ss^{\prime}} .
\end{eqnarray*}
$$  \eqno(4.7)$$
In eqs.(4.3), $b^{\dagger}_{k,s} (b_{k,s})$ are creation (annihilation)
operators for the positive - energy particles and $d_{- k,s}
(d^{\dagger}_{-k,s})$ are creation (annihilation)
operators for the negative - energy particles (anti-particles).

Connecting eqs.(4.2) and (4.4), it is obtained that
$$ \Big( i \gamma^{\mu}D_{\mu}  -  M_f  \Big) \psi_{I k,s} = 0,
 \eqno(4.8a)$$  
$$ \Big( i \gamma^{\mu}D_{\mu}  -  M_f  \Big) \psi_{II k,s} = 0, 
\eqno(4.8b)$$  

Now using the operator $ \Big(- i \gamma^{\mu}D_{\mu}  -  M_f  \Big)$
 from the
left in eq.(4.8a) as
$$\Big(- i \gamma^{\mu}D_{\mu}  -  M_f  \Big) \Big( i \gamma^{\mu}D_{\mu}  -
M_f  \Big) \psi_{I k,s} = 0, $$   
it is obtained that [17]
$$ \Big({\Box} + (4 \eta)^{-1} <{\tilde R}> + \sigma^2 <{\tilde R}>^2
\Big)\psi_{I k,s} = 0, \eqno(4.9a)$$    
where ${\tilde R} = \eta R$ and
$${\Box} = \frac{1}{\sqrt{- g}}\frac{\partial}{\partial x^{\mu}}\Big[
\sqrt{- g} g^{\mu\nu} \frac{\partial}{\partial x^{\nu}}\Big]. 
 \eqno(4.9b)$$

Similarly,  from eq.(4.7b), it is obtained that
$$ \Big({\Box} + (4 \eta)^{-1} <{\tilde R}> + \sigma^2 <{\tilde R}>^2
\Big)\psi_{II k,s} = 0. \eqno(4.10)$$

When temperature $T$ is not very much below $T_c $
one obtains the vacuum state $<{\tilde R}> = 0,$ where eqs.(4.7)
 and (4.8)
reduce to
$$ {\Box}\psi_{I k,s} = 0 = {\Box}\psi_{II k,s} \eqno(4.11a,b) $$

Particles are created due to conformal symmetry breaking, which is caused by the mass term. In the state  $<{\tilde R}> = 0,$ there is no term to break this symmetry. So, production of spin-1/2 particles are not possible in this state.

As given in  section 2, one obtains the state
 $<{\tilde R}> =
{1 \over 2} T_c$ when the temperature falls sufficiently below
 $T_c.$ In this
state, the universe obeys the geometry given by the line-element (2.31) .

In what follows, creation of spin-1/2 particles is investigated 
in the
model given by this line-element. In the background geometry of this model,
 for
every $k, l, m $ and $s,$ eq.(4.9) looks like

$$ e^{i k.r} \Big[{\ddot f_{k l m,s}} + \frac{3 {\dot a}}{a} {\dot
f_{k l m,s}} + \Big(\Big\{\frac{k^2}{a^2} + \frac{ l (l + 1)}{a^2 r^2} +
 \frac{ 6 r_0 ( 2 r - r_0 )}{7 a^2 r^2 ( r - r_0 )^2}\Big\} \Big[ 1
- \frac{r_0}{r} \Big]^{-4/7}$$ $$ + {\tilde M}^2_f \Big) f_{k l m,s} \Big]  = 0
  \eqno(4.12a)$$ 
connecting eqs.(4.4a) and (4.9). Here
$${\tilde M}^2_f = (4\eta)^{-1} <{\tilde R}> + \sigma <{\tilde R}>^2$$
$$ = \frac{T_c}{8\eta} + \frac{\sigma^2 T_c^2}{4}.   \eqno(4.12b)$$ 

Using the convolution theorem [25] as above and integrating over $r$,
 one obtains that
$$ {\int^{\infty}_{r_0 + \epsilon}}e^{i k.r} {dr}
 \Big[{\ddot f_{k l m,s}} +
\frac{3 {\dot a}}{a} {\dot 
f_{k l m,s}} + \Big\{\frac{k^2}{a^2 \eta } {\int^{\infty}_{r_0
 + \epsilon}}
e^{ - i k.y} \Big[ 1 - \frac{r_0}{y}
\Big]^{- 4/7}{dy} $$ $$+ \frac{ l (l + 1)}{a^2 \eta} 
{\int^{\infty}_{r_0 +
\epsilon}} y^{-2} \Big[ 1 - \frac{r_0}{y}
\Big]^{-4/7}e^{ - i k.y} {dy} + \frac{6r_0}{7 a^2 \eta} \times $$
 $${\int^{\infty}_{r_0 +
\epsilon}} \frac{( 2 y - r_0 )}{ y^2 ( y - r_0 )^2} \Big[ 1 
- \frac{r_0}{y} \Big]^{-4/7} e^{ - i k.y} {d y}+ {\tilde M}^2_f \Big\} f_{k
l m,s} \Big]  = 0 .  \eqno(4.13)$$ 

Details of the evaluation of integrals with respect to $y$ are
 given in
the Appendix B. Using these results, in eq.(4.13), one obtains
$${\ddot f_{k l m,s}} + \frac{3 }{2} \sqrt{\frac{T_c}{6\eta}}
 {\dot f_{k l
m,s}} + \Big[ \frac{X_{k l m}}{a^2_{10}} e^{-(t - t_{1e})\sqrt{T_C/6\eta}} +
{\tilde M}^2_f \Big] f_{k l m,s} = 0 , \eqno(4.14a)$$
where
$$X_{k l m,s} = ( 1/a^2 \eta ) \epsilon^{3/7}
 (r_0 + \epsilon)^{4/7}
Cos \{ k (r_0 + \epsilon)\} 
\Big[  k^2 + \frac{ l ( l + 1)}{ ( r_0 + \epsilon )^2}$$  $$ -
\frac{12}{7} r_0 \epsilon^{-2} (r_0 + \epsilon)^{-1} + \frac{6}{ 7} r_0^2
\epsilon^{-2} ( r_0 + \epsilon )^{-2} \Big] . \eqno(4.14b)$$ 

Eqs.(4.14) yield the solution for $t>0$,
$$f_{k l m,s} = C_1 e^{\pm \frac{3}{4} (t - t_{10})
 \sqrt{T_c/6\eta}}
J_{\pm 2 i \alpha}\Big(\gamma_{k l m, s} 
 e^{ - \frac{1}{2} (t - t_{1e})
\sqrt{T_c/6\eta}} \Big),  \eqno(4.15a)$$ 
where $C_1$ is a normalization constant,
$$\gamma^2_{k l m, s} = \frac{ 24 \eta X_{k l m, s}}{a^2_{10} T_c}
\eqno(4.15b)$$  
and
$$ \alpha^2 = \frac{6 \eta}{T_c} \Big[{\tilde M}^2_f
 - \frac{3 T_c}{32 \eta}
\Big] = \frac{6 \eta}{T_c} \Big[\frac{1}{4} \sigma^2 T_c^2 +
\frac{5T_c}{32 \eta} \Big]  \eqno(4.15c)$$  
using the definition of ${\tilde M}^2_f$ from eq.(4.12b).

Connecting eqs.(4.5a) and (4.15)

$$\psi_{I k,s} =  \Big[ (2 \pi \eta)^{1/2} r \Big( 1 - \frac{r_0}{r}
\Big)^{1/7} \Big]^{-1} {\sum_{l,m}} C_1
 e^{- \frac{3}{4} (t - t_{1e})
\sqrt{T_c/6\eta}} \times$$
$$J_{\pm 2 i \alpha}\Big(\gamma_{k l m, s} 
 e^{ - \frac{1}{2} (t - t_{1e})
\sqrt{T_c/6\eta}} \Big)  Y_{l m}(\theta, \phi)
 e^{ i k.r} u_s . \eqno(4.16)$$

Normalization of $\psi_{I k,s}$ is done using the rule (4.7c)
 at the $t = t_{1e}$ hypersurface denoted as $\Sigma$ onwards. As a result,
 $C_1$ is
obtained as 
$$ C_1 = \Big| J_{\pm 2 i \alpha}(\gamma_{k l m, s} ) \Big|^{-1}.
\eqno(4.17)$$ 
Thus
$$\psi_{I k,s} =  \Big[ (2 \pi \eta)^{1/2} r \Big( 1 - \frac{r_0}{r}
\Big)^{1/7} \Big]^{-1} {\sum_{l,m}} \Big| J_{\pm 2 i \alpha}
(\gamma_{k l
m, s} ) \Big|^{-1}  e^{- \frac{3}{4} (t - t_{1e})
\sqrt{T_c/6\eta}} \times$$
$$J_{\pm 2 i \alpha}\Big(\gamma_{k l m, s} 
 e^{ - \frac{1}{2} (t - t_{1e})
\sqrt{T_c/6\eta}} \Big)  Y_{l m}(\theta, \phi) e^{ i k.r} u_s .
 \eqno(4.18)$$
  
Similarly  
$$\psi_{II k,s} =  \Big[ (2 \pi \eta)^{1/2} r \Big( 1 - \frac{r_0}{r}
\Big)^{1/7} \Big]^{-1} {\sum_{l,m}}
 \Big| J_{\pm 2 i \alpha}(\gamma_{-k l
m, s}) \Big|^{-1}  e^{- \frac{3}{4} (t - t_{1e})
\sqrt{T_c/6\eta}} \times$$
$$J_{\pm 2 i \alpha}\Big(\gamma_{-k l m, s} 
 e^{ - \frac{1}{2} (t - t_{1e})
\sqrt{T_c/6\eta}} \Big)  Y_{l m}(\theta, \phi)
 e^{- i k.r} {\hat u}_s .
\eqno(4.19)$$ 

Asymptotics of these solutions, when $t\to \infty$, yield
$$\psi^{\rm out}_{I k,s} = 
 \Big[ (2 \pi \eta)^{1/2} r \Big( 1 - \frac{r_0}{r}
\Big)^{1/7} \Big]^{-1} {\sum_{l,m}}
 \Big| J_{\pm 2 i \alpha}(\gamma_{k l
m, s} ) \Big|^{-1}  e^{- \frac{3}{4} (t - t_{1e})
\sqrt{T_c/6\eta}} \times$$
$$ \Big[\Gamma{(1 - 2 i \alpha)}\Big]^{-1}
 \Big(\frac{\gamma_{k l m,
s}}{2}\Big)^{-2i\alpha}  e^{ + i \alpha (t - t_{1e}) 
\sqrt{T_c/6\eta}}   Y_{l m}(\theta, \phi) e^{ i k.r} u_s 
. \eqno(4.20a)$$
 and
 $$\psi^{\rm out}_{II k,s} =  \Big[ (2 \pi \eta)^{1/2} r
 \Big( 1 - \frac{r_0}{r}
\Big)^{1/7} \Big]^{-1} {\sum_{l,m}} \Big| J_{\pm 2 i \alpha}
(\gamma_{-k l
m, s} ) \Big|^{-1}  e^{- \frac{3}{4} (t - t_{1e})
\sqrt{T_c/6\eta}} \times$$
$$ \Big[\Gamma{(1 - 2 i \alpha)}\Big]^{-1}
 \Big(\frac{\gamma_{-k l m,
s}}{2}\Big)^{-2i\alpha}  e^{ + i \alpha (t - t_{1e}) 
\sqrt{T_c/6\eta}}   Y_{l m}(\theta, \phi)
 e^{- i k.r} {\hat u}_s .
\eqno(4.20b)$$ 
For $t\to - \infty$, one obtains
$$\psi^{\rm in}_{I k,s} = 
 \Big[ (2 \pi \eta)^{1/2} r \Big( 1 - \frac{r_0}{r}
\Big)^{1/7} \Big]^{-1} {\sum_{l,m}}
 \Big| J_{\pm 2 i \alpha}(\gamma_{k l
m, s}) \Big|^{-1}  e^{ \frac{3}{4} (t - t_{1e})
\sqrt{T_c/6\eta}} \times$$
$$ \Big[\Gamma{(1 + 2 i \alpha)}\Big]^{-1} 
\Big(\frac{\gamma_{k l m,
s}}{2}\Big)^{-2i\alpha}  e^{ - i \alpha (t - t_{1e}) 
\sqrt{T_c/6\eta}}   Y_{l m}(\theta, \phi) e^{ i k.r} u_s .
 \eqno(4.21a)$$
 and
 $$\psi^{\rm in}_{II k,s} = 
 \Big[ (2 \pi \eta)^{1/2} r \Big( 1 - \frac{r_0}{r}
\Big)^{1/7} \Big]^{-1} {\sum_{l,m}}
 \Big| J_{\pm 2 i \alpha}(\gamma_{-k l
m, s}) \Big|^{-1}  e^{ \frac{3}{4} (t - t_{1e})
\sqrt{T_c/6\eta}} \times$$
$$ \Big[\Gamma{(1 + 2 i \alpha)}\Big]^{-1}
\Big({\frac{\gamma_{-klm,s}}{2}}\Big)^{-2i\alpha} 
 e^{ - i \alpha (t -
t_{1e})\sqrt{T_c/6\eta}}  Y_{l m}(\theta, \phi) e^{- i k.r} {\hat
u}_s . \eqno(4.21b)$$  
In eqs.(4.20) - (4.21), $\Gamma{(x)}$ stands for the gamma function
 of $x.$

In the in- as well as out- region,

the decomposed form of $\psi$
 can be
written as [4]
\begin{eqnarray*}
 \psi &=& {\sum_{s=\pm1}}{\sum_k} \Big( b_{k,s}^{\rm in}
 \psi^{\rm in}_{I
(k,s)} + d^{\dagger {\rm in}}_{-k,-s} \psi^{\rm in}_{II (-k,-s)}
 \Big)\\
&=& {\sum_{s=\pm1}}{\sum_k} \Big( b_{k,s}^{\rm out}
 \psi^{\rm out}_{I 
(k,s)} + d^{\dagger{\rm out}}_{-k,-s}
 \psi^{\rm out}_{II (-k,-s)} \Big),
\end{eqnarray*}
$$ \eqno(4.22a,b)$$ 
as both in- and out-spinors belong to the same Hilbert space. The in- and
out-vacuum states are defined as 
$$  b_{k,s}^{\rm in}|{\rm in} > = d_{-k,s}^{\rm in}|{\rm in} > = 0
\eqno(4.23a,b)$$  
and
$$  b_{k,s}^{\rm out}|{\rm out} > = d_{-k,s}^{\rm out}|{\rm out} > = 0
\eqno(4.23c,d)$$

Bogoliubov transformations are given as [3,4]
\begin{eqnarray*}
b_{k,s}^{\rm out} & = & b_{k,s}^{\rm in} \alpha_{k,s} +
d_{-k,-s}^{{\dagger}{\rm in}}\beta_{k,s} 
 \\ b_{k,s}^{{\dagger}{\rm out}} &
= &  \alpha^*_{k,s} b_{k,s}^{{\dagger}{\rm in}} +
 \beta^*_{k,s} d_{-k,-s}^{\rm 
in}  \\ d_{-k,-s}^{{\dagger}{\rm out}} & = &
 b_{k,s}^{\rm in} \alpha_{k,s} +
d_{-k,-s}^{{\dagger}{\rm in}}\beta_{k,s}  \\ 
 d_{-k,-s}^{\rm out} &
= &  \alpha^*_{k,s} b_{k,s}^{{\dagger}{\rm in}} +
 \beta^*_{k,s} d_{-k,-s}^{\rm 
in} .
\end{eqnarray*}
$$ \eqno(4.24a,b,c,d)$$
Connecting eqs.(2.22)-(2.24), one obtains [22]
$$|\alpha_{k}|^2 + |\beta_{k}|^2 = \sum_{s}|\alpha_{k.s}|^2 +
|\beta_{k.s}|^2 = 1 ,  \eqno(4.25)$$
where
$$\alpha_{k.s} = {\int_{\Sigma}} \sqrt{ - g_{\Sigma}} {d^3x} {\bar
\psi}_{I(k,s)}^{\rm out} {\tilde \gamma}^0 \psi_{I(k,s)}^{\rm in}
\eqno(4.26a)$$ 
and
$$\beta_{k.s} = {\int_{\Sigma}} \sqrt{ - g_{\Sigma}} {d^3x} {\bar
\psi}_{II(k,s)}^{\rm out} {\tilde \gamma}^0 \psi_{II(k,s)}^{\rm in}
\eqno(4.26b)$$ 

Connecting eqs.(4.20a), (4.21a) and (4.26a)
\begin{eqnarray*}
\alpha_{k.s} &=& (1/2)\Big[\Gamma{(1 - 2 i \alpha)}\Big]^{-2}
\Big(\frac{\gamma_{k l m, s}}{2}\Big)^{-4i\alpha} \Big| J_{\pm 2 i
\alpha}(\gamma_{k l m, s} ) \Big|^{-2} \\ & =& \frac{sinh (2 \pi \alpha)} {(4
\pi \alpha)}\Big[\Gamma{(1 - 2 i \alpha)}\Big]^{-2}.
\end{eqnarray*}
 $$ \eqno(4.27a)$$ 
Similarly eqs.(4.20b), (4.21b) and (4.26b)
\begin{eqnarray*}
\beta_{k.s} &=& (1/2)\Big[\Gamma{(1 + 2 i \alpha)}\Big]^{-2}
\Big(\frac{\gamma_{k l m, s}}{2}\Big)^{-4i\alpha} \Big| J_{\pm 2 i
\alpha}(\gamma_{k l m, s} ) \Big|^{-2} \\ & =& \frac{sinh (2 \pi \alpha)} {(4
\pi \alpha)}\Big[\Gamma{(1 + 2 i \alpha)}\Big]^{-2}.
\end{eqnarray*}
 $$ \eqno(4.27b)$$

Conditions (4.25) and eqs.(4.27) yield an useful result
 $$ |\alpha_{k.s}|^2 = |\beta_{k.s}|^2 = \frac{1}{4} \eqno(4.28)$$
suggesting that $\alpha$ should be very small. 
using the symmetry $k \to -k$ caused by symmetry of the model under
 $r \to -r.$

The relative probability of creation of a particle - antiparticle
 pair is given as 
$$\omega_{k l m,s} = \Big| \beta_{k.s}/\alpha_{k.s} \Big|^2 = 1.
\eqno(4.29)$$  

Absolute probability of the creation of particle - antiparticle
 pairs
requires the total probability of creating $0,1,2, \cdots$ pairs
 to be
unity [5], which means that
$$N_{k l m,s}(1 + \omega_{k l m,s} + \omega_{k l m,s}^2 + \cdots)
 = 1, $$
where $N_{k l m,s}$ is the probability of creation of no pair of
particle-antiparticle. Using eq.(4.29), it is obtained that
$$N_{k l m,s} = \frac{1}{1 + 1 + 1 + \cdots} = 0. $$
It shows that probability of vacuum to remain vacuum is
$$|<{\rm out}| {\rm in}>|^2 = {\prod_{k l m,s}}N_{k l m,s} = 0.
\eqno(4.30)$$     
implying certainity of creation of particle-antiparticle pairs.

The action $S$, given by eq.(2.11), yields components of energy - momentum tensor for $\psi$ field  as [22,27]
\begin{eqnarray*}
T_{\mu\nu} &=& \frac{1}{\sqrt{-g}}
 \frac{\delta S}{\delta g^{\mu\nu}} \\ &=& i{\bar \psi}  \gamma_{\mu} D_{\nu}
 \psi - \sigma \eta R_{\mu\nu}{\bar \psi}\psi - {1 \over 2}
g_{\mu\nu} {\bar \psi} \Big( i  \gamma^{\rho} D_{\rho}  - \sigma {\tilde R}  \Big)
 \psi \\&& - \sigma \eta ( D_{\mu} D_{\nu} - g_{\mu\nu} {\Box} ){\bar \psi}\psi + c.c. 
\end{eqnarray*}
$$ \eqno(4.31)$$

Connecting eqs.(3.20a) and (4.31), energy density of created particles in the state ${\tilde R} = 1/2 T_c$ is obtained as
\begin{eqnarray*}
\rho = < T^0_0 > & = & < {\rm in}|- i {\bar \psi}\gamma^0 \partial_0   -
\sigma \eta R^0_0 {\bar \psi}\psi + 2 \sigma {\tilde R}{\bar \psi}\psi +
c.c.|{\rm in}>\\&& - < {\rm out}|- i {\bar \psi}\gamma^0 \partial_0   - \sigma
\eta R^0_0 {\bar \psi}\psi + 2 \sigma {\tilde R}{\bar \psi}\psi + c.c.|{\rm
  out}> \\ &=& < {\rm in}|- i {\bar \psi}\gamma^0 \partial_0   + 1/2 \sigma
T_c{\bar \psi}\psi + c.c.|{\rm in}> \\&& - < {\rm out}|- i {\bar \psi}\gamma^0 \partial_0    + 1/2 \sigma T_c{\bar \psi}\psi + c.c.|{\rm out}>
\end{eqnarray*}
$$ \eqno(4.32)$$ 
using $R^0_0 = 3 {\ddot a}/a = T_c / {2 \eta}.$

In- and out- $\psi$, given by eqs.(4.20) and (4.21), recast eq.(4.32) as

$$ \rho = \sum_{klm}\frac{8 \alpha}{\eta} \frac{\epsilon^{8/7}}{(r_0 + \epsilon)^{22/7}}
cos^2 [k(r_0 + \epsilon)] sinh[(3/2)(t - t_{1e}) \sqrt{T_c/6\eta}] \eqno(4.33)$$
using convolution theorem for $\psi$ as given in the preceding section.

Similarly trace of stress - tensor for created spin-1/2 particles is given as

$$ < T > = <{\rm in}| T_{\mu\nu} |{\rm in} > - <{\rm out}| T_{\mu\nu} |{\rm
  out} > = 0  , \eqno(4.34)$$
using In- and out- solutions for $\psi$, given by eqs.(4.20) and (4.21).

Eq.(4.28) shows number density  of created spin-1/2 particles as

$$|\beta_k|^2 = \frac{1}{2} \eqno(4.35)$$
yielding energy density of created particles also as

$$ \rho = \frac{1}{2} M_f . \eqno(4.36)$$
  $ |\beta_k|^2 $ gives number of created particles per unit volume during the time period
  $(t_{20}- t_{10}) = 1.38 \times 10^7 t_P$, so evaluating $\rho$, from eq.(4.32) and comparing with
  the same from eq.(4.36), it is obtained that

$$ M_f \simeq 54.87 \times 10^{99}  \frac{\epsilon^{8/7}}{(r_0 +
  \epsilon)^{22/7}}.   \eqno(4.37)$$

 Here also summation is done with the help of the Riemann
zeta function defined in the earlier section.

\vspace{0.5cm}

\centerline{\bf 4. Conclusions}

\bigskip

Thus, in this part of the series, production of spinless and spin-1/2
particles are discussed in the first two phases of the universe. Contribution
of these particles to later development of the universe to third and fourth
phases of the universe will be discussed in paper II.

\bigskip

\centerline{\bf Acknowledgement}

Author is thankful to Prof.K.P.Sinha for helpful suggestions and
encouragement.
\vspace{0.5cm}

\bigskip

\centerline{\bf Appendix A}

\centerline{\underline{\bf Riccion and Graviton}}

\smallskip

From the action

$$ S = \int {d^4 x} {d^D y}  \sqrt{- g_{(4+D)}} \quad
\Big[ \frac{M^{(2+D)}R_{(4+D)}}{16 \pi
} + {\alpha_{(4+D)}} R_{(4+D)}^2 + $$
$$\gamma_{(4+D)} ( R_{(4+D)}^3   - \frac{6(D+3)}{D-2)}{\Box}_{(D+4)}R^2_{(D+4)})\Big], \eqno(A.1)$$
the gravitational equations are obtained as

$$ \frac{M^{(2+D)}}{16 \pi} (R_{MN} - {1 \over 2} g_{MN} R_{(4+D)} ) + {\alpha_{(4+D)}} H^{(1)}_{MN} + {\gamma_{(4+D)}} H^{(2)}_{MN} = 0,\eqno(A.2a)$$ 
where

$$ H^{(1)}_{MN} = 2 R_{; MN} - 2 g_{MN} {\Box}_{(4+D)} R_{(4+D)} - {1\over 2} g_{MN} R^2_{(4+D)} + 2 R_{(4+D)} R_{MN}, \eqno(A.2b)$$  
and
$$ H^{(2)}_{MN} = 3 R^2_{; MN} - 3 g_{MN} {\Box}_{(4+D)} R^2_{(4+D)} - \frac{6(D+3)}{(D-2)}\{- { 1 
\over 2} g_{MN}{\Box}_{(4+D)} R^2_{(4+D)}$$  $$+ 2{\Box}_{(4+D)}R_{(D+4)}R_{MN} + R^2_{; MN}\} - { 1 \over 2} g_{MN} R^3_{(4+D)} + 3 R^2_{(4+D)} R_{MN} . \eqno(A.2c) $$

Taking $g_{MN} = \eta_{MN} + h_{MN}$ with $\eta_{MN}$ being $(4+D)$-dimensional Minkowskian metric tensor components and $h_{MN}$ as small fluctuations, the equation for $graviton$ are obtained as

$${\Box}_{(4+D)} h_{MN} = 0   \eqno(A.3) $$
neglecting  higher-orders of $h$.

On compactification of $M^4 \otimes S^D$ to $M^4$, eq.(A.3) reduces to the equation for 4-dimensional $graviton$ as
$${\Box} h_{\mu\nu} + \frac{l(l+D-1)}{\rho^2}h_{\mu\nu} = 0   \eqno(A.4) $$
for the space time 

$$ d S^2 = g_{\mu\nu} d x^{\mu} d x^{\nu} - \rho^2 [ d\theta_1^2 + sin^2\theta_1 d\theta^2_2 + \cdots +
sin^2\theta_1 \cdots sin^2\theta_{(D-1)} d\theta^2_D.  \eqno(A.5)$$

The 4-dimensional graviton equation (A.4) is like usual 4-dimensional graviton equation ( the equation derived from 4-dimensional action) only for $l = 0$. Thus the massless graviton is obtained for $l = 0$ only.

As explained, in section 2, the trace of equations (A.2) leads to the $riccion$ equation

$$[{\Box} + {1 \over2} \xi R + m^2_{\tilde R} + \frac{\lambda}{3!} {\tilde R}^2]{\tilde R} + \vartheta  = 0,  \eqno(A.6a)$$
where

\begin{eqnarray*}
 \xi & = & \frac{D}{2(D+3)} +  \eta^2 \lambda R_D \\ m^2_{\tilde R} & = & - \frac{(D + 2) \lambda V_D}{16 \pi G_{(4+D)}} + \frac{D R_D}{2(D+3)} + \frac{1}{2} \eta^2 \lambda R_D^2 \\ \lambda  & = & \frac{1}{4(D+3)\alpha}, \\ \vartheta & = & - \eta \Big[- \frac{(D + 
2) \lambda M^{(2+D)}V_D}{16 \pi} + \frac{D  R_D^2}{4 (D+3)} + \frac{1}{6}
\eta^2 \lambda R_D^3 \Big].
\end{eqnarray*}

The graviton $h_{\mu\nu}$ has 5 degrees of freedom (2 spin-2 graviton, 2
spin-1 gravi-vector (gravi-photon) and 1 scalar). The scalar mode $f$
satisfies the equation

$$ {\Box} f + \frac{l(l + D - 1)}{\rho^2} f = 0 \eqno(A.7)$$
Comparison of eqs.(A.6) and (A.7) show many differences between scalar mode
$f$ of graviton and the riccion $(\tilde R)$ e.g. $\xi, \lambda$ and
$\vartheta$ , given by eqs.(A.6b,c,d,e), are vanishing for $f$, but
non-vanishing for $\tilde R$. Eq.(A.7) shows $(mass)^2$ for $f$ as 

$$ m_f^2 = \frac{l(l + D - 1)}{\rho^2} f , \eqno(A.8)$$
whereas $(mass)^2$ for $\tilde R$, given by eq.(A.6c), depends on $G_{4+D},
V_D $ and $R_D$ (given in section 2). $m_f^2 = 0$ for $l = 0,$ but $
m^2_{\tilde R}$ can vanish only when gravity is probed upto $\sim 10^{-33}
cm.$ As mentioned above, so far, gravity is probed only upto 1cm. Thus
$(mass)^2$ of riccion does not vanish. 

So, even though, $f$ and ${\tilde R}$ are scalars arising from gravity, both
are different. $Riccion$ can not arise without higher-derivative curvature
terms in the gravitational action, but $graviton$ can be obtained even from
Einstein-Hilbert action.

$${\Box} = \frac{1}{\sqrt{-g}} \frac{\partial}{\partial x^{\mu}}
\Big(\sqrt{-g} g^{\mu\nu} \frac{\partial}{\partial x^{\nu}} \Big) =
\eta^{\mu\nu}\frac{\partial}{\partial X^{\mu}\partial X^{\nu}  }, $$
where ${X^{\mu}}$ are locally inertial co-ordinates and $\eta^{\mu\nu}$ are
Minkowskian metric components. It shows that the scalar like operator ${\Box}$
has the same role on $\tilde R$ as it is for other scalar fields $\phi$ due to
principle of equivalence. 

\centerline{\bf Appendix B}

\bigskip
In eq.(3.6), using the convolution theorem the function $e^{i k.r} \Big[1
- \frac{r_0}{r} \Big]^{-4/7}$ can be written as
$$e^{i k.r} \Big[1 - \frac{r_0}{r} \Big]^{-4/7} = \eta^{-1} e^{i k.r}
{\int_{r_0 + \epsilon}^{\infty}}{d y} e^{-i k.y} \Big[1 - \frac{r_0}{y}
\Big]^{-4/7} , \eqno(B.1)$$ 
where $\eta$ has dimension of length, as used in section 2,for dimensional
correction. Here in the integral $d y$ introduces a quantity of length
dimension which is compensated by $\eta.$

Evaluation of the integral in eq.(B.1), is done as follows.
\begin{eqnarray*}
{\int_{r_0 + \epsilon}^{\infty}}{d y} \frac{e^{-i k.y}}{ \Big[1 - \frac{r_0}{y}
\Big]^{4/7}} &=& {\int_{r_0 + \epsilon}^{\infty}}{d y} \frac{y^{4/7} (y -
r_0)^{\omega}}{(y - r_0)^{\omega + 4/7} } {\sum_{n =0}^{\infty}} \frac{( -
i k y)^n}{n!}\\ &=&{\sum_{n =0}^{\infty}} \frac{( -i k )^n}{n!}
\frac{1}{\Gamma(\omega + 4/7)} {\int_{r_0 + \epsilon}^{\infty}} {d y} y^{n +
4/7} (y - r_0)^{\omega}\times \\&& \Big[ 
{\int_0^{\infty}} x^{\omega - 3/7} e^{- x (y - r_0)} {d x} \Big] 
\\ &=& {\sum_{n =0}^{\infty}} \frac{( -i k )^n}{n!} \frac{1}{\Gamma(\omega
+ 4/7)} 
{\int_0^{\infty}} {d x} x^{\omega - 3/7} e^{ x r_0)} \times \\&& \Big[{\int_{r_0 +
\epsilon}^{\infty}}  y^{n + 4/7} (y - r_0)^{\omega} e^{- x y} {d y} \Big]
 \\& \simeq& {\sum_{n =0}^{\infty}} \frac{( -i k )^n}{n!}
\frac{1}{\Gamma(\omega + 4/7)} 
{\int_0^{\infty}} {d x}  x^{\omega - 3/7} e^{- x \epsilon} \times \\&& (r_0 +
\epsilon)^{(n + 4/7)} \epsilon^{\omega} \Big[\frac{1}{x} +
\frac{\omega}{\epsilon x^2}  + \frac{\omega (\omega - 1)}{\epsilon^2  x^3}
+ \cdots \Big] 
\end{eqnarray*}
$$ \eqno(B.2)$$
neglecting terms containing $(r_0 + \epsilon)^{-1}$ compared to terms
containing $ \epsilon^{-1}.$

Performing further integration in eq.(B.2), one obtains that
\begin{eqnarray*}
{\int_{r_0 + \epsilon}^{\infty}}{d y} \frac{e^{-i k.y}}{ \Big[1 - \frac{r_0}{y}
\Big]^{4/7}} & \simeq& {\sum_{n =0}^{\infty}}\frac{( -i k )^n}{n!}
\frac{1}{\Gamma(\omega + 4/7)} (r_0 + \epsilon)^{(n + 4/7)}\times \\&&
\epsilon^{3/7} 
\Big[\Gamma(\omega - 3/7) + \omega \Gamma(\omega - 10/7) + \\&& 
 \omega (\omega - 1) \Gamma(\omega - 17/7) + \cdots \Big]
 \\&=& {\sum_{n =0}^{\infty}}\frac{( -i k )^n}{n!}
\frac{1}{\Gamma(\omega + 4/7)} (r_0 + \epsilon)^{(n + 4/7)} \epsilon^{3/7}
\\&=& (r_0 + \epsilon)^{ 4/7} \epsilon^{3/7} cos k(r_0 +
\epsilon),
\end{eqnarray*}
$$ \eqno(B.3)$$  
when $\omega \to \infty$. Here, symmetry under $k \to -k$ caused by
symmetry of the model under $r \to -r$ is used.

The same procedure yields
$${\int_{r_0 + \epsilon}^{\infty}}{d y} e^{-i k.y} y^{-2} \Big[1 -
\frac{r_0}{y} \Big]^{-4/7} = (r_0 + \epsilon)^{-10/7} \epsilon^{3/7} cos
k(r_0 + \epsilon) ,  \eqno(B.4)$$

$${\int_{r_0 + \epsilon}^{\infty}}{d y} e^{-i k.y} y^{-3/7} [y - r_0]^{-18/7}
 = (r_0 + \epsilon)^{-3/7} \epsilon^{-11/7} cos
k(r_0 + \epsilon) ,  \eqno(B.5)$$

$${\int_{r_0 + \epsilon}^{\infty}}{d y} e^{i k.y} y^{-4/7} [y - r_0]^{4/7}
 = (r_0 + \epsilon)^{-4/7} \epsilon^{11/7} cosk(r_0 + \epsilon) ,  \eqno(B.6)$$
 
$${\int_{r_0 + \epsilon}^{\infty}}{d y} e^{-i k.y} y^{-10/7} [y - r_0]^{-18/7}
 = (r_0 + \epsilon)^{-10/7} \epsilon^{-11/7}cosk(r_0 + \epsilon) .
\eqno(B.7)$$

\centerline{\bf REFERENCES}
\vspace{0.5cm}

\noindent1. K.S.Stelle; Phys.Rev.D, {\bf 16},(1977) 923.

\bigskip

\noindent2. V.Tr. Gurovich and A.A.Starobinsky; Zh. Eksp.Theor. Fiz. {\bf 77} (1977) 1683 (Sov.Phys.JETP, {\bf 50(5)} (1979)844).

\bigskip

\noindent3. B.S. DeWitt; Phys.Rev., {\bf 160},(1967) 113.

\bigskip

\noindent4. Ya. B. Zel'dovich and I.D.Novikov ;Relyativistskaya Astrofizica ( Relativistic Astrophysics), Fizmatgiz (1968).

\bigskip

\noindent5. Ya. B. Zel'dovich and L.P.Pitaevski; Comm. Math. Phys., {\bf 23} (1971) 185.

\bigskip

\noindent6. A.A.Starobinsky; Phys.Lett., {\bf 91B} (1980) 99.

\bigskip

\noindent7. L.A.Kofmann, A.D.Linde and A.A.Starobinski; Phys.Lett., {\bf 157B} (1985) 361.

\bigskip

\noindent8. M.B.Mijic, M.S.Morris and Wai-Mo Suen; Phys.Rev.D, {\bf 34} (1986) 2934.

\bigskip

\noindent9.  Wai-Mo Suen and P.R.Anderson; Phys.Rev.D, {\bf 15(10)} (1987) 2940.

\bigskip

\noindent10. B.Whitt; Phys. Lett. {\bf 145 B} (1984)176; S.W.Hawking $\&$ J.C.Luttrell , Nucl.Phys.B, {\bf 247} (1984) 250 ; N.H.Barth $\&$ S.M.Christensen, Phys.Rev.D, {\bf 28} (1983) 1876 .

\bigskip

\noindent11.  S.K.Srivastava and K.P.Sinha ; Phys. Lett. B{\bf 307} (1993)
40 ; J. Ind. Math. Soc., {\bf 61} (1994) 80.
\bigskip

\noindent12.  K.P.Sinha and S.K.Srivastava ; Pramana {\bf 44} (1995) 333.
\bigskip

\noindent13.  S.K.Srivastava and K.P.Sinha ; Int. J. Theor. Phys. {\bf 35}
(1996) 155.
\bigskip

\noindent14. S.K.Srivastava and K.P.Sinha ; Mod. Phys. Lett. A {\bf 38}
(1997) 2933.
\bigskip

\noindent15. S.K.Srivastava ; IL Nuvo Cimento {\bf 113} (1998) 1239. 
\bigskip

\noindent16. S.K.Srivastava ; Int. J. Mod. Phys. A {\bf 14} (1999) 875.

\bigskip

\noindent17. S.K.Srivastava ; Mod.Phys.Lett. A. {\bf 14} (1999) 1021.
\bigskip

\noindent18.  P.D.B.Collins, A.D.Martin and E.J.Squires ; Particle Physics
and Cosmology , Wiley , New York , 1989.
\bigskip

\noindent19. S.Weinberg ; Gravitation and Cosmology, John Wiley $\&$ Sons
(1972).

\bigskip

\noindent20.A.B. Lahnas, N.E. Mavromatos and D.V. Nanopoulos,  Int. J. Mod. Phys. D, {\bf 12(9)}, 1529 (2003).

\bigskip

\noindent21.  S.W.Hawking and G.F.R.Ellis ; The large scale structure of
space-time , Cambridge University Press  (1973), 150.
\bigskip

\noindent22. S.K.Srivastava and K.P.Sinha ; Aspects of Gravitational
Interaction, Nova Science Publications, New York (1998).

\bigskip

\noindent23.  A.H.Guth ; Phys. Rev. D{\bf 22} (1981) 347.

\bigskip

\noindent24.A. Riess $et$ $al.$, Astron. J. {\bf 116}, 1009
(1998); astro-ph/9805201; S. J. Perlmutter $et$ $al.$, Astrophys. J. {\bf
  517}, 565 (1999); astro-ph/9812133; J.L. Tonry $et$ $al.$,
astro-ph/0305008;  D. N. Spergel $et$ $al.$,  astro-ph/0302209;  J. A. Peacock $et$ $al.$, Nature, {\bf 410}, 169 (2001).

\bigskip

\noindent25. J.Mathews and R.L.Walker ; Mathematical Methods of Physics,
Benjamin, Inc. Massachusetts (1970).

\bigskip

\noindent26. N.D.Birrell and P.C.W. Davies ; Quantum Fields in Curved
Spaces, Cambridge University Press, Cambridge (1982). 

\bigskip

\noindent27. B.S.DeWitt ; Physics Reports  {\bf 19} (1975) 295.

\bigskip

\noindent28. E. Mottola ; Phys.Rev.D {\bf 31} (1985) 754.

\end{document}